# Evidence of heavy fermion physics in the thermoelectric transport of magic angle twisted bilayer graphene

Rafael Luque Merino[1,2,3], Dumitru Călugăru[4], Haoyu Hu[5], Jaime Díez-Mérida[1,2,3], Andrés Díez-Carlón[1,2,3], Takashi Taniguchi[6], Kenji Watanabe[7], Paul Seifert[1,8], B. Andrei Bernevig[4,5,9], Dmitri K. Efetov[2,3*]

1. ICFO - Institut de Ciencies Fotoniques, The Barcelona Institute of Science and Technology, Castelldefels, Barcelona, 08860, Spain
2. Fakultät für Physik, Ludwig-Maximilians-Universität, Schellingstrasse 4, 80799 München, Germany
3. Munich Center for Quantum Science and Technology (MCQST), München, Germany
4. Department of Physics, Princeton University, Princeton, NJ 08544, USA
5. Donostia International Physics Center (DIPC), Paseo Manuel de Lardizabal. 20018, San Sebastián, Spain
6. International Center for Materials Nanoarchitectonics, National Institute for Materials Science, 1-1 Namiki, Tsukuba 305-0044, Japan
7. Research Center for Functional Materials, National Institute for Materials Science, 1-1 Namiki, Tsukuba 305-0044, Japan
8. Institute of Physics, Faculty of Electrical Engineering and Information Technology, University of the Bundeswehr Munich and SENS Research Center, Werner-Heisenberg-Weg 39, 85577 Neubiberg, Germany
9. IKERBASQUE, Basque Foundation for Science, 48013 Bilbao, Spain

*E-mail: dmitri.efetov@lmu.de

**It has been recently postulated, that the strongly correlated flat bands of magic-angle twisted bilayer graphene (MATBG) can host coexisting heavy and light carriers. While transport and spectroscopic measurements have shown hints of this behavior, a more direct experimental proof is still lacking. Here, we explore the thermoelectric response of MATBG through the photo-thermoelectric (PTE) effect in gate-defined MATBG *pn*-junctions. At low temperatures, we observe sign-preserving, filling-dependent oscillations of the Seebeck coefficient at non-zero integer fillings of the moiré lattice, which suggest the preponderance of one carrier type despite tuning the Fermi level from hole to electron doping of the correlated insulator. Furthermore, at higher temperatures, the thermoelectric response provides distinct evidence of the strong electron correlations in the unordered, normal state. We show that our observations are naturally accounted for by the interplay of light and long-lived and heavy and short-lived electron bands near the Fermi level at non-zero integer fillings. Our observations firmly establish the electron and hole asymmetry of the correlated gaps in MATBG, and shows excellent qualitative agreement with the recently developed topological heavy fermion model (THF).**

The flat bands of magic-angle twisted bilayer graphene[1] (MATBG) offer a rich playground for condensed matter physics, as they host both strong electronic interactions and non-trivial topology[2]. A multitude of symmetry-broken ground states emerge at non-zero integer fillings of the moiré unit cell $\nu$, where $\nu = 4n/n_s$ and $n_s$ is the carrier density in a fully-occupied moiré band. The contrasting nature of its ground states (which include superconductivity[3–5], Mott-like physics[4,6] and topological states[7–10]) suggest the coexistence of itinerant and localized electrons within the flat bands. In addition, the sawtooth-like evolution of the electronic entropy[11–15] near integer $\nu$, as well as the appearance of Landau fans only away from charge neutrality[3–6], points to a pronounced asymmetry of the charge $\pm 1$ excitations of the correlated ground states.

This apparent dichotomy has motivated the development of 'heavy fermion' models of MATBG[16–21]. There, localized, heavy electron bands account for the strongly-correlated phenomena while itinerant, highly dispersive electron bands govern the system's metallicity and non-trivial topology. Heavy fermion models provide a unified framework for MATBG phenomenology, yet experimental evidence of distinct transport properties, such as strong lifetime asymmetry between light and heavy electrons, remains limited. For this purpose, Seebeck coefficient measurements can be of particular interest as these can accurately probe asymmetries in the mass and mobility of the carriers below and above the Fermi level, as has been demonstrated in semimetals and narrow gap semiconductors, where both electrons and holes participate in transport[22,23].

In this work, we explore the low temperature thermoelectric response of the flat bands in MATBG through optical excitation of gate-defined $pn$-junctions. We find unambiguous evidence of photo-induced thermovoltage generation driven by the Seebeck effect in the MATBG flat bands. The photo-thermoelectric (PTE) response develops oscillations at each integer filling, which arise from the formation of correlated insulating states. Remarkably, the thermoelectric response maintains an electron-like character throughout the conduction flat band, despite the reconstructing Fermi surface. This points to a scenario where transport is dominated by light electrons, which have much larger mean free path than the heavy holes which would otherwise give rise to a hole-like, positive thermoelectric response. We provide a natural interpretation of these observations via the topological heavy fermion (THF) picture of MATBG[16,29,30], where the distinct transport properties of light and heavy electrons inherently account for the observed response. Furthermore, we shed light on the high-temperature thermoelectricity of the flat bands, highlighting the persistent influence of strong electron correlations even in the absence of an ordered state.

We probe the low-temperature thermoelectric response of the flat bands using laser excitation ($\lambda = 1550$ nm) to locally heat up a MATBG $pn$-junction, as sketched in Fig. 1a. We focus on two high-quality MATBG samples ($\theta = 1.14°$ and $\theta = 1.06°$) that exhibit archetypal correlated resistive states at non-zero integer fillings $\nu$ (Fig. 1b), as well as superconducting phases (see SI). A gate-defined $pn$-junction is created by splitting the top graphite gate, allowing independent control of the carrier concentration on each side. We define said junction in a section of the Hall bar device with a homogeneous twist angle near 1.1°. Figure 1c depicts the dual top-gate $R_{XX}$ map of the junction, where both sides of the junction feature pronounced correlated states which can be addressed independently. The fabrication of these samples is detailed in the Methods and previous work[31].

The continuous-wave excitation induces a local increase of the electronic temperature $\Delta T_e$ centered at the junction's interface. Using the split top gates, we establish a chemical potential difference $\Delta\mu$ (and so a Seebeck coefficient difference $\Delta S$) across the junction. The gate-dependent response around the CNP exhibits multiple sign changes in a characteristic 6-fold pattern (Fig. 1e), as has been observed previously in single layer graphene devices[32]. Such a pattern is a consequence of the antisymmetric gate-dependence of $S$ in particle-hole symmetric (zero-gap) semiconductors, as the response does not only depend on the polarity of the junction ($pn$ or $np$), like for the photo-voltaic effect, but is set by the Seebeck coefficient on each side of the junction (sign of the response is reversed between $pp^+$ and $p^+p$). This clearly establishes the (photo-) thermoelectric (PTE) as the origin of the measured response[32,33].

In the SI we diligently discuss and rule out all other potential mechanisms, and derive all the assumptions about the thermal mechanism governing MATBG.

We consistently find such 6-fold symmetric response around CNP across multiple samples and various experimental conditions (Fig. S18). This results in a zero bias, net photo-thermoelectric voltage, which can be approximated by $V_{PTE} = -(S_2 - S_1)\Delta T_e$ where $S_i(\mu_i)$ denotes the Seebeck coefficient of each side of the junction[32–35] (see lower left sketch in Fig. 1a). We note that the PTE effect confers multiple advantages for the study of thermoelectric transport, including the spatial control of the $\Delta T_e$ profile and the efficient carrier heating in graphene-like systems[36,37].

The observed PTE response also indicates graphene-like, hot carrier dynamics in MATBG for above-gap excitation. Here, the absorbed photon energy is efficiently converted into an increased temperature $T_e + \Delta T_e$ of the carrier distribution at $E_F$, while the phononic degrees of freedom stay in equilibrium at lattice temperature $T_L$[38]. As the MATBG flat bands lie near $E_F$, thermoelectric transport from the hot carriers serves as a powerful tool to probe the low-energy electronic spectrum of MATBG. This picture is consistent with the existing studies of light-matter interaction in MATBG in this wavelength range[39,40]. In the following, we leverage the photo-thermoelectric response of the *pn*-junction $V_{PTE}$ to investigate the low-energy electronic spectrum of the strongly-interacting flat bands.

We further explore the thermoelectric response of the *pn*-junction, concentrating on the conduction flat band ($\nu > 0$) as it features stronger correlated states in transport. In Figure 1d, the dual gate map of the PTE response $V_{PTE}(\nu_1, \nu_2)$ reveals multiple sign changes appearing at integer $\nu_i$ for each side of the junction. These features coincide with the correlated states in the dual gate $R_{xx}$ map, indicating that the measured photo-response captures the interaction-driven gaped states along the flat band[11–13].

To study these oscillations in more detail the measurement scheme is simplified by fixing one side of the junction at CNP, where $S_2 = S_{CNP} = 0$. Then, the PTE response just reads $V_{PTE} = -(S_2 - S_1)\Delta T_e = S_1 \Delta T_e$, enabling the direct mapping of the Seebeck coefficient $S_1$. This simplified scheme resembles the conventional configuration used in Joule heating approaches that feature a region of homogeneous $S$ and an asymmetric temperature gradient $\Delta T_e$. We restrict ourselves to the linear heating regime ($\Delta T_e \ll T_e = T_L$) by using low excitation powers. The electronic temperature increase $\Delta T_e$ is estimated through a two-temperature model that includes the experimentally determined thermal relaxation time of the devices (SI Section IV).

The low temperature, gate-dependent thermoelectric response $V_{PTE}(\nu_1)$ is shown in Fig. 2a for both samples, where the absorbed optical power for devices 1 and 2 is $P_{abs}$ = 2.12 µW and $P_{abs}$ = 0.93 µW, respectively. We find that the response across the CNP is conventional, featuring a symmetric, sign-changing doping dependence, where holes give rise to a positive $S$ and electrons to a negative $S$. Upon further doping of the flat bands, we observe oscillatory features of the thermoelectric response near each integer $\nu$. These oscillations can be attributed to the formation of symmetry-broken ground states[29,30]. Strikingly, the $V_{PTE}$ remains electron-like (negative) and is highly asymmetric for $\nu = +1, +2$, despite the prominent gap-like response at those fillings (Figs. S9, S10). This asymmetric sign-preserving behavior contrasts with the conventional semiclassical picture, where the Seebeck coefficient $S_{Mott}$ is expected to change sign across the energy gap (extrema of $R_{xx}(\nu_1)$), as the carrier type changes. This anomalous thermoelectric response points to a pronounced asymmetry of the charge excitations around $E_F$.

In Figs. 2b-d we explore possible sources for the particle-hole asymmetry observed in the MATBG electron-doped correlated insulators. The first scenario, known as the 'Dirac revival' picture, postulates that the degenerate flat bands can be modeled as flavor-polarized Dirac cones[11]. This minimal model was previously used to interpret non-semiclassical thermoelectricity in the MATBG flat bands[25]. Upon reaching an integer filling, all the spectral weight is shifted onto a single Dirac cone which falls below $E_F$. Consequently, an asymmetry in the density of states $N(E)$ appears: $N(E)$ is greater for holes as electron-like excitations lie near the Dirac point, where $N(E)$ is minimized. The Seebeck coefficient would then be expected to display excess hole-like contributions (Fig. 2b), contrary to or observations. We note that in this picture (which neglects correlation effects) the asymmetry in $\sigma(E)$ originates solely from the asymmetry of $N(E)$, as the lifetime of the carriers $\tau(E)$ and their group velocity $v_g(E)$ are constant within the aforementioned 'Dirac bands'.

Upon the interaction-induced reconstruction of its Fermi surface, the MATBG flat bands are known to host correlated insulating ground states at integer fillings $\mathbb{Z}$. The charge-one excitations above these ground states can be phenomenologically modelled by two quadratically-dispersing bands around $E_F$ (Fig. 2c)[41,42]. We initially contemplate strong mass asymmetry of the bands as the potential source for the pronounced *e-h* asymmetry. We consider a scenario where the hole band has larger effective mass $m_h^* > m_e^*$, while all carriers share a common carrier lifetime $\tau(E)$. We use a two-band model[29] to compute the Seebeck coefficient under conditions of extreme mass asymmetry ($m_h^* = 150\ m_e^*$). The Seebeck coefficient (lower panel of Fig. 2c) exhibits small excess electron-like contributions but still features a zero-crossing across the gap. The persistence of the sign change indicates the need for comprehensive modeling of the correlated insulators to account for the experimental observations, as effective mass asymmetry by itself is not enough to explain the observed behavior.

Hence, realistic modelling of the transport properties of the MATBG correlated insulators must account for the energy (or band) dependence of the carrier lifetime $\tau(E)$. We now include a band-dependent carrier lifetime into the two-band model (Fig. 2d). We observe electron-like thermoelectricity if we consider $\tau_h < \tau_e$, while keeping the mass asymmetry ratio fixed at $m_h^* = 150\ m_e^*$. In this scenario, hole-like transport is suppressed due to the combined effects of the increased scattering rate, which reduces $\tau_h$, and a lower group velocity $v_g(E)$ of the hole band. The computed $S$ for $\tau_e = 6\tau_h$ is depicted in the lower panel of Fig. 2c. Notably, the Seebeck coefficient exhibits electron-like characteristics for integer filling $\nu = \mathbb{Z}$, even for significant hole doping of the correlated insulator. This observation strongly suggests that $\tau(E)$, whose energy (band) dependence is often overlooked, strongly influences the thermoelectric response of such general correlated insulator state. Modified transport coefficients due to strong asymmetry of $\tau(E)$ have been reported in other strongly-correlated systems[43,44], including heavy fermion compounds[45]. In localized heavy bands, *e-e* interactions introduced a large scattering rate and, therefore, a small lifetime for the associated carriers, diminishing their transport contribution.

While the above toy model already contains the main clues to explain the observed anomalous PTE responses - the mass and lifetime asymmetry - we further compare our results with the recently developed topological heavy fermion model (THF) of MATBG[16]. In the THF model, the flat bands arise from the hybridization between strongly-correlated, highly localized

'heavy' f-electrons and strongly-dispersive, 'light' c-electrons. Colloquially, the topology is carried by the light electrons while the flatness of the bands is a direct consequence of the localized natured of the heavy electrons. Therefore, this model naturally incorporates coexisting electronic species with contrasting transport coefficients. The c-electrons form coherent excitations that dominate the transport properties, while the incoherent excitations formed by f-electrons do not contribute directly to transport. Indirectly, however, the strong interactions between localized f-electrons can break the global symmetries of MATBG, inducing gaps in the dispersion of the c-electrons.

Using self-consistent 2$^{nd}$ order perturbation theory[29,30], we compute the interacting energy bands and corresponding Seebeck coefficients for the correlated insulating ground states of MATBG at a twist angle $\theta = 1.06°$. All computed band structures and $S$ correspond to $T \sim 0.6\text{-}0.7 \, T_{order}$, where $T_{order}$ denotes the self-consistently determined ordering temperature of each ground state. The low-temperature band structures of the Kramers intervalley coherent (KIVC)[16,46], or KIVC + valley polarized (VP), ground states near integer fillings $\nu = 0, +1, +2$ are presented in Figs. 3a, 3d, 3g, respectively. Color coding denotes the light (c) or heavy (f) character of the band. Generally, the c-electron states make up the bands near $\Gamma$ while f-states appear away from the Brillouin zone center. The essential properties of the THF band structures, we discuss here, do not depend on the choice of the ground state[29]. Other states, such as the intervalley coherent Kekulé spiral state[47], can be considered but do not change the light-heavy dichotomy of the dispersion around correlated insulators. At $\nu = 0$, the bands near $E_F$ consist of 'light' c-electron states for both holes and electrons. As the charge $\pm 1$ excitations are symmetric around $E_F$, the Seebeck coefficient (Fig. 3b) exhibits a conventional antisymmetric lineshape, in agreement with our observations for the thermoelectric response of MATBG near CNP (Fig. 3c).

The interacting THF band structures for the ground states appearing at non-zero integer fillings (Fig. 3d, 3g) exhibit a marked asymmetry of the charge $\pm 1$ excitations. We find that the low-energy hole excitations correspond to localized f-electrons, while itinerant c-electrons form the bands just above $E_F$. The strong interactions between f-electrons lead to a reduced (increased) carrier lifetime (scattering rate) for the 'heavy' electrons near $E_F$. Indeed, we find that the correlated insulators at positive, non-zero integer fillings realize the 'light-heavy' scenario sketched in Fig. 3d, as the weakly-dispersive hole band features states with low carrier lifetime $\tau$.

We then compute the Seebeck coefficient for the correlated insulators, whose (doped) band structures are illustrated in Figs. 3d, 3g. The hole-like contributions to thermoelectricity are quenched, due to the reduction of $\tau$ for the localized f-states, leading to a fully negative Seebeck coefficient across the interaction-driven gaps near $\nu = +1, +2$ (Figs. 3e, 3h). These findings provide excellent agreement with the experimentally observed thermoelectricity (Figs. 3f, 3i). The THF band structures and computed Seebeck coefficient for different ground states, fillings and temperatures are presented in Ref. 29. We note that for $\nu = +1, +2$ the negative peak of the Seebeck coefficient is shifted towards CNP. This effect, which is reproduced in our theoretical calculations, stems from the highly broadened spectral weight of the f-electron bands below $E_F$[33].

Overall, the 'light-heavy' structure of the correlated insulators within the THF model provides a natural interpretation for the observed PTE response. We stress the crucial role of finite carrier lifetime associated with the localized f-electrons in the observation of a negative, sign-preserving Seebeck coefficient. From the experimental data and the estimated $\Delta T_e$, we find $S \sim 50 - 150$ μV/K for $T_L = 10$ K across the conduction flat band; in good agreement with the theoretically computations[29] and previous reports[25–27]. We note that whereas for most semimetals $S$ is reduced due to electron-hole compensation, here the band asymmetry arising from the 'light-heavy' structure of the correlated insulators leads to a large Seebeck coefficient.

We now turn our focus to the thermoelectric response at higher lattice temperatures. As the temperature rises, the symmetry-broken ground states disappear and MATBG transitions into a symmetry-preserving, unordered state[48], with Hubbard bands seen in STM experiments[49,50]. Theoretical calculations estimate critical temperature for this symmetric state around 10-20 K. The PTE response at $T_L = 30$ and 50 K using a $P_{abs} = 5.47$ μW is illustrated in Fig. 4a. In contrast to the low-temperature behavior, we observe hole-like thermoelectricity with $S > 0$ for $0 < \nu < 3$, which we attribute to the temperature activated carriers in the higher energy bands. The overall trend of the signal mimics the thermoelectric response in the non-interacting limit (gray trace in Fig. 4a). However, distinct gap-like oscillations appear around each integer $\nu$ (Fig. 4b), underscoring the existence of persistent electron correlations in the symmetric (unordered) state.

To model the high-temperature thermoelectric response, we employ the same THF model with identical parameters used for the symmetry-broken calculations. However, we now examine symmetric solutions, in which none of the model's symmetries are spontaneously broken[29,30]. The computed Seebeck coefficient $S_{Sym}$ for $T_L = 15$ K is illustrated in Fig. 4c. $S_{Sym}$ exhibits marked oscillations at integer $\nu$ along with a positive (hole-like) offset, which can be traced back to the thermoelectric response in a non-interacting scenario[29]. The oscillations arise from the opening of gaps (few meV) through interactions between the localized f-states. Overall, the THF model can qualitatively reproduce the observed thermoelectric response at high temperatures under the only assumption of a symmetry-preserving solution. The temperature mismatch between theory and experiment stems from less precise modeling of the non-interacting dispersion, which can be significantly altered by extrinsic effects, such as strain or superlattice relaxation.

The measured PTE response at elevated temperatures highlights the presence of sizable electron correlations beyond the ordering temperatures of the symmetry-broken ground states. The qualitative match with the Seebeck coefficient for symmetric solution of the THF model provides further evidence of 'heavy fermion' physics in MATBG. These findings also stress that electron interactions can induce gap openings in the electronic spectrum of the flat band even when all the symmetries of the system are preserved.

To summarize, we have established the PTE effect as the photovoltage generation mechanism in MATBG for above-gap excitation. Taken at face-value, our observations unveil 'heavy fermion' physics in MATBG, which underscores the natural description of the system using two distinct carrier types, with both distinct masses and distinct scattering times. Our study stimulates the further investigation of Kondo and heavy fermion phenomena in 2D materials[51–54]. From an application standpoint, the large, gate-tunable thermoelectric response of MATBG also opens the door for twist angle engineering of thermoelectric devices[55]. broadband absorption[56], efficient carrier heating[36,37] and ultrafast thermal relaxation[57].

## Methods

**Device fabrication:** The MATBG devices were fabricated using a cut and stack technique. All flakes were first exfoliated on a Si/SiO$_2$ (285 nm) substrate and later picked up using a polycarbonate (PC)/polydimethylsiloxane (PDMS) stamp. All the layers were picked up at a temperature of $\sim 100$ °C. We used an AFM tip to cut the graphene in order to avoid strain during the pick-up process. The PC/PDMS stamp picks up first the top graphite layer, the top hBN and the first graphene layer. Before picking up the second graphene layer, we rotate the stage by an angle of $\theta = 1.1°$. Finally, the stamp picks up the bottom hBN and bottom graphite gates. We drop the finalized stack on a Si/SiO$_2$ substrate by melting the PC at $\sim 180$ °C. The resulting stack is etched into a Hall bar using a CHF$_3$/O$_2$ plasma and a 1D contacts are formed by evaporating Cr (5 nm)/Au (50 nm), see Fig. S1b. We etch a narrow channel of $\sim 150\ nm$ in the top gate using an O$_2$ plasma. Before etching the top gate, the device was characterized in transport using a 4-probe configuration at $T_L = 35\ mK$ to identify the pair of contacts closest to the magic angle of $\theta \approx 1.1°$. The junction was made in between this pair of contacts.

**Transport measurements:** Transport studies for the characterization of the 2 samples were carried out in a dilution refrigerator (BlueFors SD250) with a base temperature of

20 mK and a VTI cryostat (ICEOxford) with a base temperature of 1.55 K. Further transport measurements were done *in situ* in the optical cryostat (Attodry 800, base temperature 6 K) used for the optoelectronic measurements. All transport measurements are performed using a standard low-frequency lock-in technique (Stanford Research SR860 amplifiers) with frequency $f = 17.177 \, Hz$.

**Optoelectronic measurements:** We study the optoelectronic response of the MATBG pn-junctions using standard DC and low-frequency AC transport measurements combined with scanning laser microscopy. All optoelectronic measurements are performed in an Attodry 800 cryostat with free-space optical access. Further information on the optoelectronic setup can be found in the SI.

## Acknowledgments

We thank N. Regnault, A. Jaoui, E. Y. Andrei, S. Buhler-Paschen, F. Koppens, L. Lin, A. Georges, A. Millis, G. Sangiovanni for useful discussions. D.C. acknowledges the hospitality of the Donostia International Physics Center, at which this work was carried out. B.A.B. was supported by DOE Grant No. DE-SC0016239. D.C. was supported by the European Research Council (ERC) under the European Union's Horizon 2020 research and innovation program (grant agreement no. 101020833) and by the Simons Investigator Grant No. 404513, the Gordon and Betty Moore Foundation through Grant No. GBMF8685 towards the Princeton theory program, the Gordon and Betty Moore Foundation's EPiQS Initiative (Grant No. GBMF11070), Office of Naval Research (ONR Grant No. N00014-20-1-2303), Global Collaborative Network Grant at Princeton University, BSF Israel US foundation No. 2018226, NSF-MERSEC (Grant No. MERSEC DMR 2011750). H.H. was supported by the European Research Council (ERC) under the European Union's Horizon 2020 research and innovation program (Grant Agreement No. 101020833) and the Schmidt Fund Grant. P.S. acknowledges support from the Alexander von-Humboldt Foundation and the German Federal Ministry for Education and Research through the Feodor-Lynen program. J.D.M. acknowledges support from the INPhINIT 'la Caixa' Foundation (ID 100010434) fellowship program (LCF/BQ/DI19/11730021). D.K.E. acknowledges funding from the European Research Council (ERC) under the European Union's Horizon 2020 research and innovation program (grant agreement No. 852927) and the German Research Foundation (DFG) under the priority program SPP2244 (project No. 535146365). K.W. and T.T. acknowledge support from the Elemental Strategy Initiative conducted by the MEXT, Japan (Grant Number JPMXP0112101001) and JSPS KAKENHI (Grant Numbers 19H05790, 20H00354 and 21H05233).

## Author contributions

R.L.M., P. S. and D.K.E conceived and designed the experiments; R.L.M., P.S., J.D.M. and A.D.C. performed the transport measurements; R.L.M. and P.S. performed the optoelectronic measurements; J.D.M and A.D.C. fabricated the samples; R.L.M. and P.S. performed the data analysis; D.C., H.H. and B.A.B. performed the theoretical modelling; T.T. and K.W. provided materials; R.L.M. and D.K.E. wrote the paper with insight from D.C., H.H. and B.A.B.

## Competing financial and non-financial interests:

The authors declare no competing financial and non-financial interests.

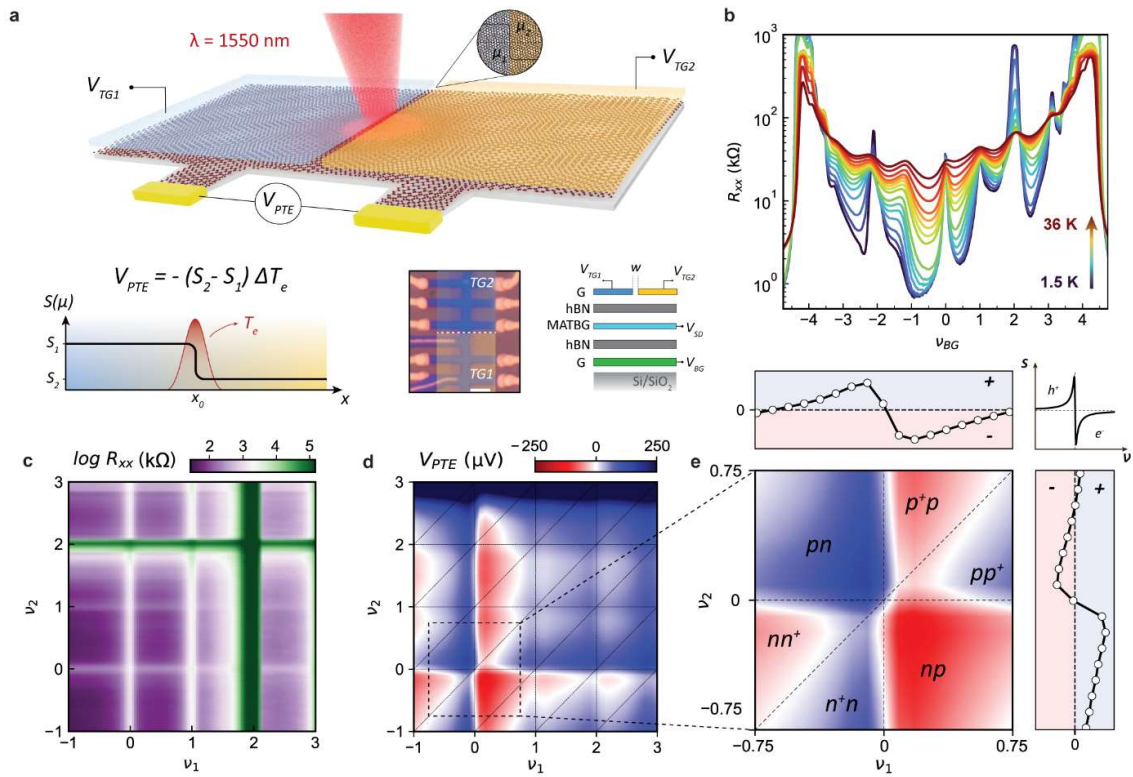

**Figure 1. Photo-thermoelectric effect in a MATBG pn-junction.** a) A local thermal gradient is created at the interface of a MATBG pn-junction using a focused laser beam. The gate-defined step of the chemical potential $\mu$ enables the generation of a net thermoelectric voltage $V_{PTE}$ across the junction. Subfigures, from left to right, depict the origin of the photo-thermoelectric voltage ($x_0$ denotes the junction position), an optical micrograph of the sample and a schematic cross-section of the heterostructure ($V_{TG}, V_{BG}$ are the top-gate and back-gate voltages, $V_{SD}$ indicates the source-drain voltage for transport measurements). b) Temperature-dependent 4-terminal resistance of Device 1 ($\theta = 1.14°$) before splitting the top gate. c) Dual-gate map of the longitudinal resistance of Device 1 at $T_L = 35$ mK. Correlated states for each side of the junction appear at integer values of $\nu_1$ and $\nu_2$. d) $V_{PTE}$ response in the electron-doped flat bands of Device 1 at $T_L = 10\ K$. Oscillations emerge around each integer filling. e) The $V_{PTE}$ response near charge-neutrality shows a characteristic 6-fold symmetry, which confirms the thermoelectric origin of the photo response.

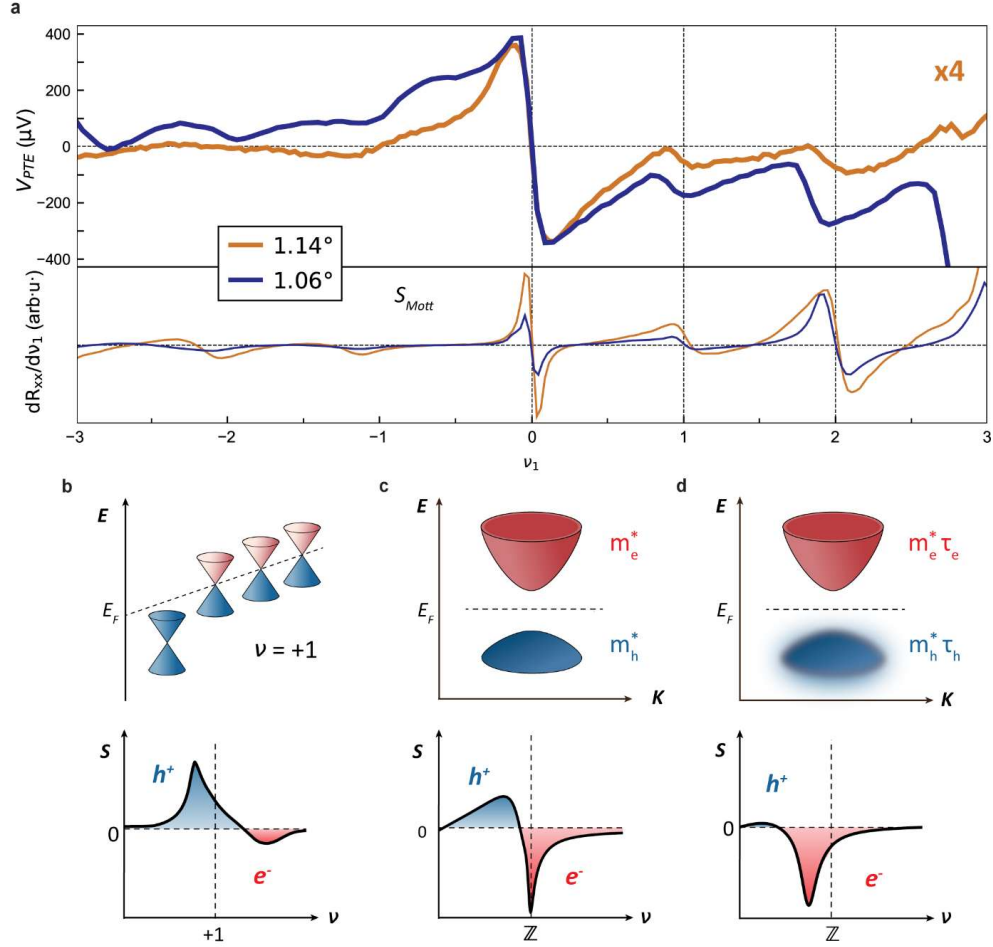

**Figure 2. Anomalous, sign-preserving thermoelectric response in the flat bands.** a) Photo-thermoelectric response across the flat bands at $T_L = 10\,K$ under low-power excitation. Lower panel shows the expectation from the semiclassical Mott formula, where $S \propto dR_{xx}/d\nu_1$. b) Upper panel: Energy diagram in the 'Dirac revival' picture for the ground state at $\nu = +1$. Lower panel: Sketch of the Seebeck coefficient for $\nu = +1$ in the 'Dirac revival' scenario, where the emergent asymmetry in the density of states leads to an excess hole-like Seebeck coefficient for $0 < \nu < 4$. c) Upper panel: Simplified two-band model for a general correlated insulator at $\nu = \mathbb{Z}$ with large effective mass asymmetry ($m_h^* \gg m_e^*$). Lower panel: Computed Seebeck coefficient in the two-band model for $m_h^* = 150\,m_e^*$. The extreme mass asymmetry leads to slight electron-like Seebeck coefficient for $\nu = \mathbb{Z} - \delta\nu$. d) Upper panel: Band structure for $\nu = \mathbb{Z}$ in a two-band, 'light-heavy' scenario where the two bands are composed by distinct electronic species with markedly different carrier lifetimes. The low carrier lifetime for the hole band is illustrated by large energy-broadening of its dispersion, as $\tau \propto 1/\Delta E$. Lower panel: Computed Seebeck coefficient for $m_h^* = 150\,m_e^*$ and $\tau_e = 6\tau_h$. The lifetime asymmetry strongly modifies the Seebeck coefficient, which evolves towards a fully negative behavior across the gap.

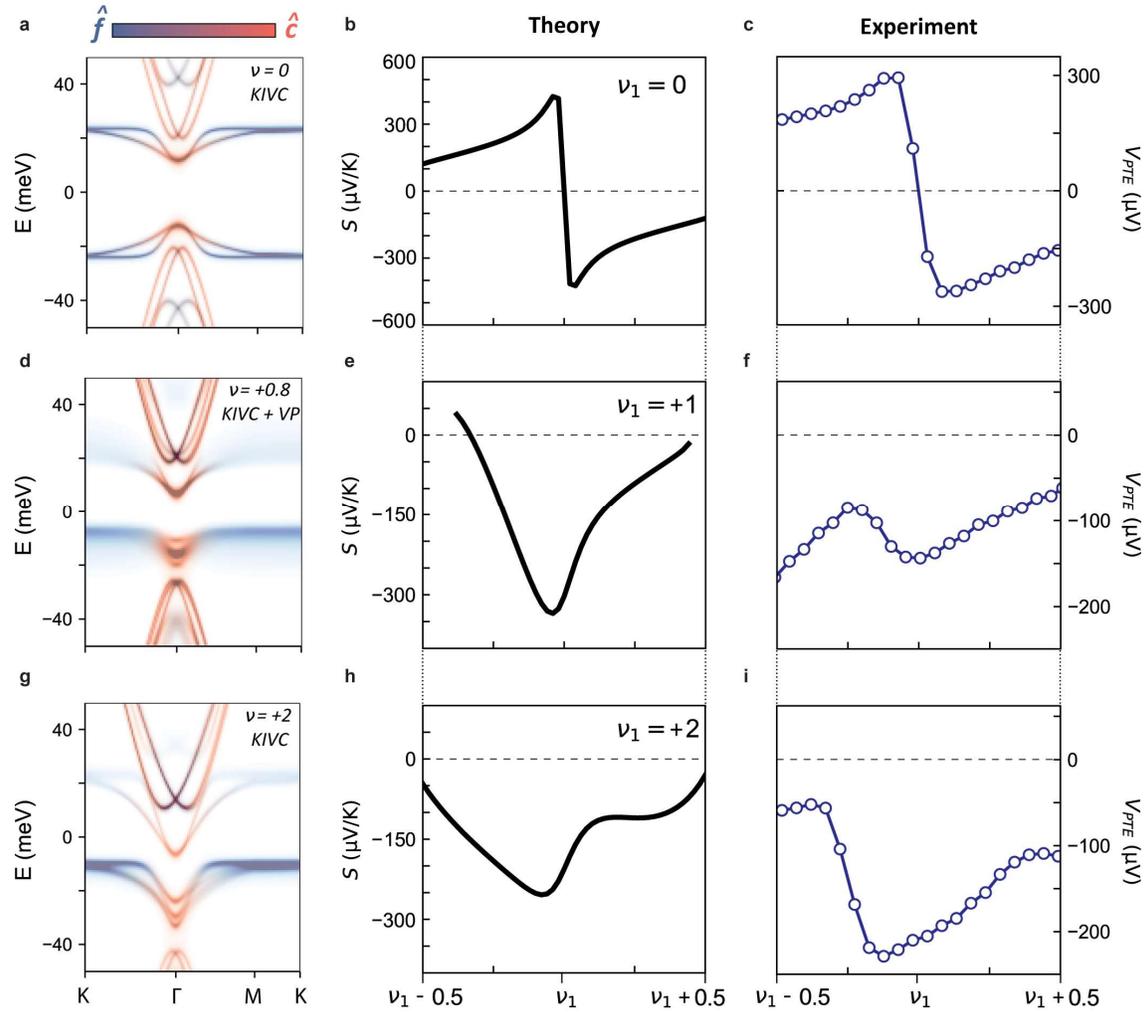

**Figure 3. Seebeck coefficient of the correlated insulators in the Topological Heavy Fermion model and comparison with experiment.** a-c) THF band structure for the KIVC correlated insulator at $\nu = 0$. Color coding of the bands indicates their light or heavy character. b) Computed Seebeck coefficient for the KIVC correlated insulator at $\nu = 0$. c) Thermoelectric response $V_{PTE} = S_1 \Delta T_e$ in the vicinity of $\nu = 0$ at $T_L = 10\ K$. d-f) THF band structure for the KIVC+VP ground state at $\nu = +0.8$. The band structure, which changes dynamically with doping, is shown away from $\nu = +1$ to highlight the spectrally broadened f-states. The localized f-electron states make up the hole band near $E_F$. e) Computed Seebeck coefficient for the KIVC correlated insulator at $\nu = +1$. f) Thermoelectric response in the vicinity of $\nu = +1$ at $T_L = 10\ K$. g-i) THF band structure for the KIVC+VP ground state at $\nu = 2$. h) Computed Seebeck coefficient for the KIVC correlated insulator at $\nu = +2$. i) Thermoelectric response near $\nu = +2$ at $T_L = 10\ K$. The measured thermoelectric response exhibits stronger oscillations for $\nu = +2$ than for $\nu = +1$.

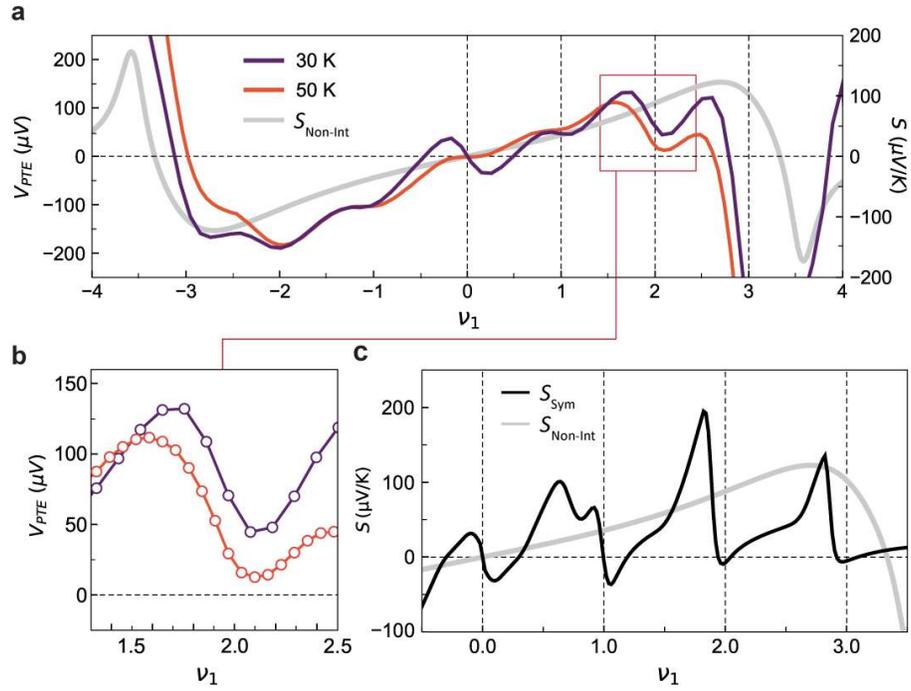

**Figure 4. Thermoelectric response in the high-temperature, symmetric state of MATBG.** a) Photo-thermoelectric response in Device 2 at $T_L = 30\,K$ and $50\,K$. The grey trace shows the expectation from the non-interacting limit of the THF model for $T_L = 15\,K$, $\tau_e/\tau_h = 4$. b) Zoom-in into the high-temperature oscillations of the thermoelectric response near $\nu = +2$. c) Seebeck coefficient computed for the symmetric solution of the THF model at $T_L = 15\,K$. Gray trace depicts the non-interacting THF Seebeck coefficient at the same temperature.

# Evidence of heavy fermion physics in the thermoelectric transport of magic angle twisted bilayer graphene

Rafael Luque Merino[1,2,3], Dumitru Călugăru[4], Haoyu Hu[5], Jaime Díez-Mérida[1,2,3], Andrés Díez-Carlón[1,2,3], Takashi Taniguchi[6], Kenji Watanabe[7], Paul Seifert[1,8], B. Andrei Bernevig[4,5,9], Dmitri K. Efetov[2,3*]

1. ICFO - Institut de Ciencies Fotoniques, The Barcelona Institute of Science and Technology, Castelldefels, Barcelona, 08860, Spain
2. Fakultät für Physik, Ludwig-Maximilians-Universität, Schellingstrasse 4, 80799 München, Germany
3. Munich Center for Quantum Science and Technology (MCQST), München, Germany
4. Department of Physics, Princeton University, Princeton, NJ 08544, USA
5. Donostia International Physics Center (DIPC), Paseo Manuel de Lardizabal. 20018, San Sebastián, Spain
6. International Center for Materials Nanoarchitectonics, National Institute for Materials Science, 1-1 Namiki, Tsukuba 305-0044, Japan
7. Research Center for Functional Materials, National Institute for Materials Science, 1-1 Namiki, Tsukuba 305-0044, Japan
8. Institute of Physics, Faculty of Electrical Engineering and Information Technology, University of the Bundeswehr Munich and SENS Research Center, Werner-Heisenberg-Weg 39, 85577 Neubiberg, Germany
9. IKERBASQUE, Basque Foundation for Science, 48013 Bilbao, Spain

*E-mail: dmitri.efetov@lmu.de

## Contents

## I. Device fabrication

Figures S1a and S1b show the two studied devices after contact evaporation (before top gate cutting). The white circles indicate the contacts used for 4-terminal transport measurements. Also indicated are the contacts to the 2 parts of the split top-gate, labelled as TG1 and TG2. The white dashed line indicates the position of the nanoscale gap separating the top gates. The graphite back gate contacts are labelled as BG. The Si-gate is grounded.

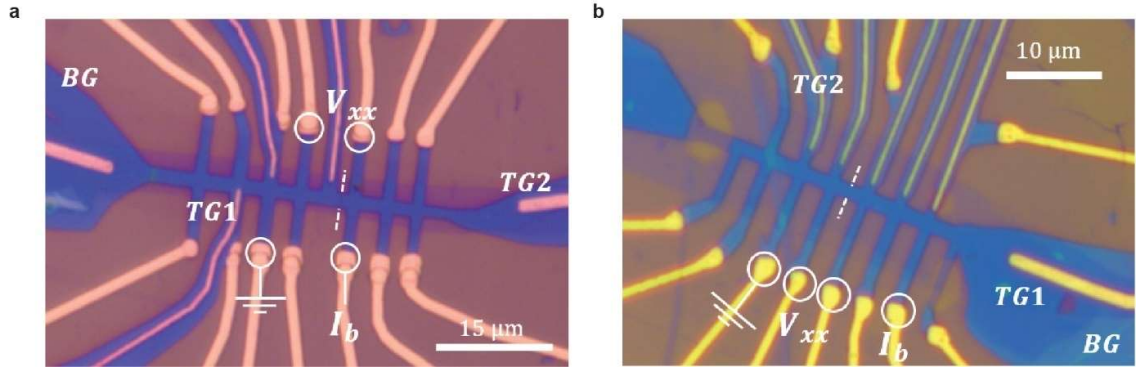

**Figure S1. Optical images of the devices.** a) Optical image for Device 1 ($\theta = 1.14°$). b) Optical image for Device 2 ($\theta = 1.06°$).

## II. Transport characterization

### a. Twist angle determination

We extract the twist angle before etching the top gate using the Landau level spectrum of the longitudinal resistance $R_{xx}$. The Landau levels emerging from the band insulators are traced back to their zero-field origin in order to better estimate the carrier density of the full filling of the moiré band $n_s$. Then, the twist angle $\theta$ can be deduced by applying the relation $n_s = \frac{8\theta}{\sqrt{3}a^2}$, where $a = 0.246$ nm is the graphene lattice constant. For Device 1 we find $n_s = (3.1 \pm 0.1) \, x 10^{12}$ cm$^{-2}$ and a twist angle of $\theta = 1.14° \pm 0.02°$. For Device 2, we use the Landau levels emerging from half-filling and obtain $n_s = (2.6 \pm 0.1) \, x 10^{12}$ cm$^{-2}$, from which we estimate a twist angle $\theta = 1.06° \pm 0.02°$.

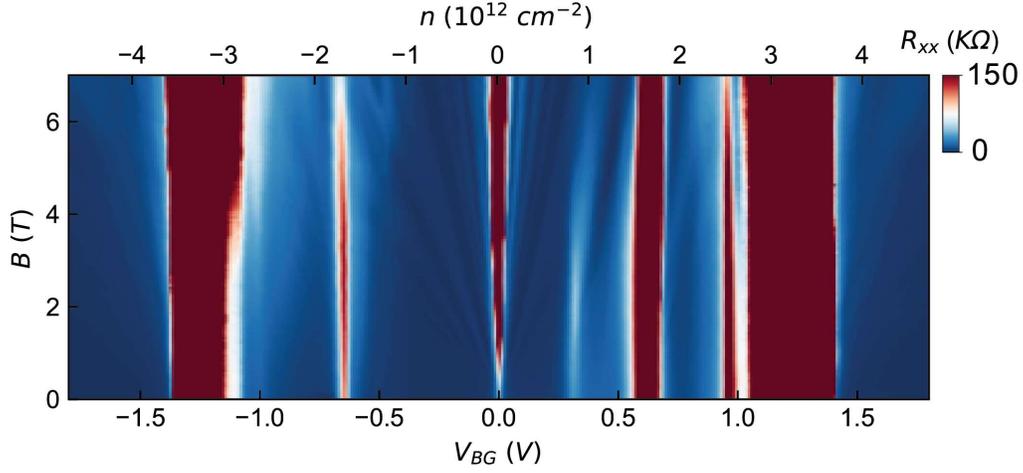

**Figure S2. Landau fan diagram for Device 1 at $T_L = 1.55$ K.** Measurement performed before cutting the top graphite gate.

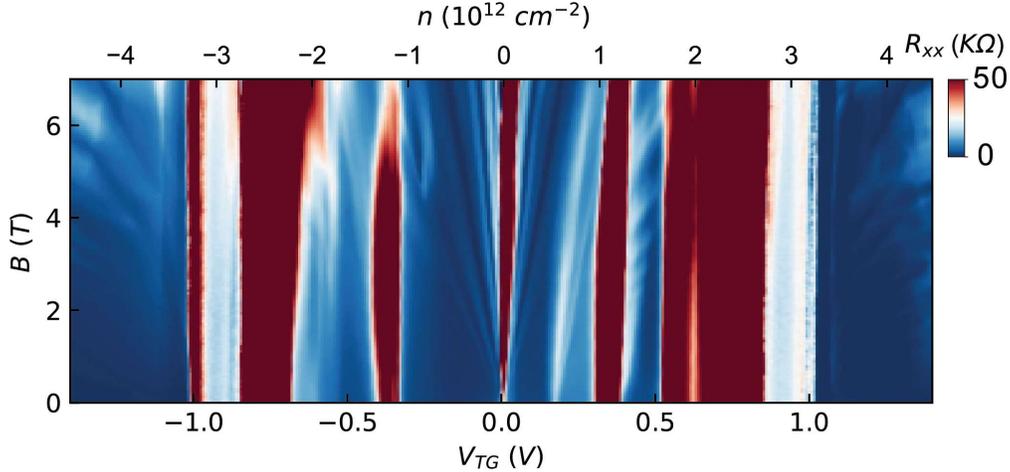

**Figure S3. Landau fan diagram for Device 2 at $T_L = 1.55$ K.** Measurement performed before cutting the top graphite gate.

## b. Superconducting states

Both devices reported in this study exhibit superconducting states near $\nu = -2$. For Device 1 ($\theta = 1.14°$), the strongest superconducting phase appears at $n = -1.79 \times 10^{-12}$ cm$^{-2}$, with a critical temperature of $T_C = 2.8$ K (Fig S4a). For Device 2 ($\theta = 1.06°$), the optimal doping for superconductivity is $n = -1.56 \times 10^{-12}$ cm$^{-2}$, where we find $T_C = 1.6$ K (Fig S5a). The current-voltage (*I-V*) characteristics, shown in Fig. S6b and S7b, exhibit strong nonlinear behavior below $T_C$ (Fig S6b, S7b).

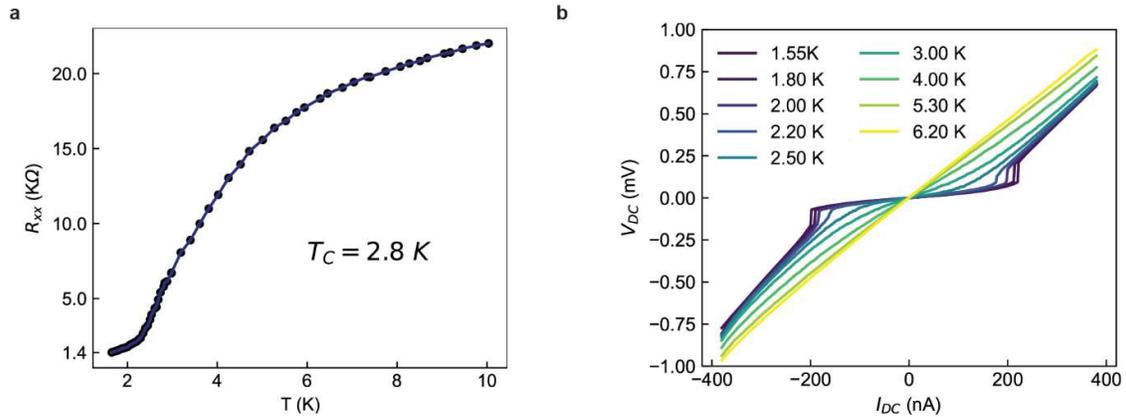

**Figure S4. Superconductivity in Device 1.** a) Superconducting transition for Device 1. b) Temperature-dependent *I-V* characteristic for Device 1 for $T_L \sim T_C$.

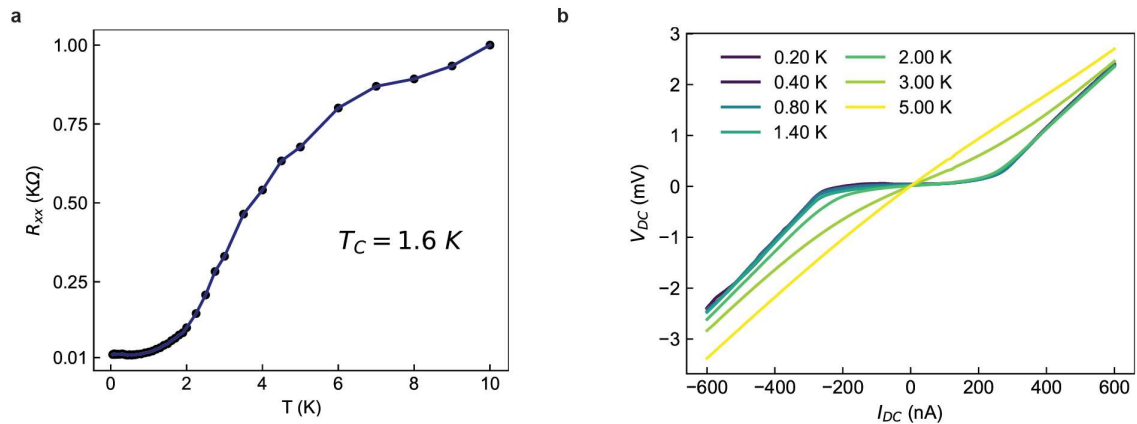

**Figure S5. Superconductivity in Device 2.** a) Superconducting transition for Device 2. b) Temperature-dependent *I-V* characteristic for Device 2 for $T_L \sim T_C$.

### c. Temperature-dependent transport, thermal activation gaps

The temperature-dependent transport properties of both devices are summarized in Figures S6-S8. The two devices exhibit comparable transport characteristics, such as strong insulating behavior for $\nu = \pm 2$, non-gapped semi-metallic characteristics for $\nu = \pm 1$ and weakly insulating charge-neutrality points. The correlated state at $\nu = \pm 3$ also exhibits insulating behavior.

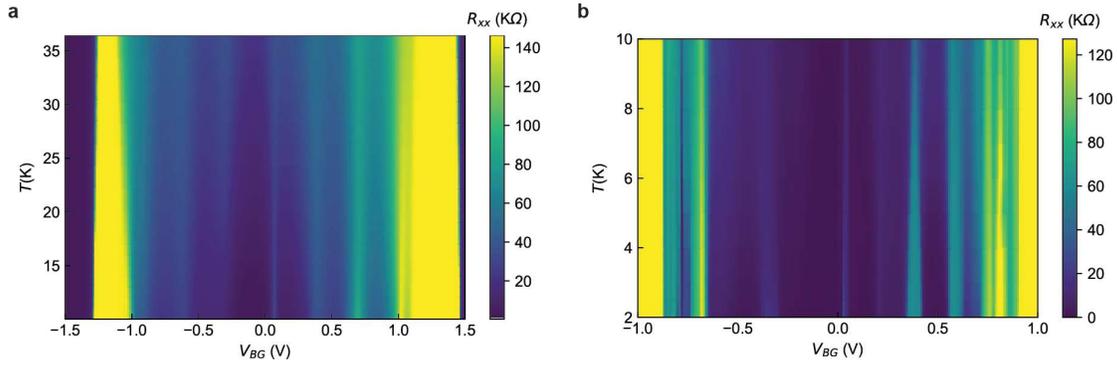

**Figure S6. Temperature-dependent transport.** a) 4-terminal $R_{xx}$ for Device 1 between $T_L = 10$ K and $T_L = 36$ K. b) 4-terminal $R_{xx}$ for Device 2 between $T_L = 1.55$ K and $T_L = 10$ K.

.

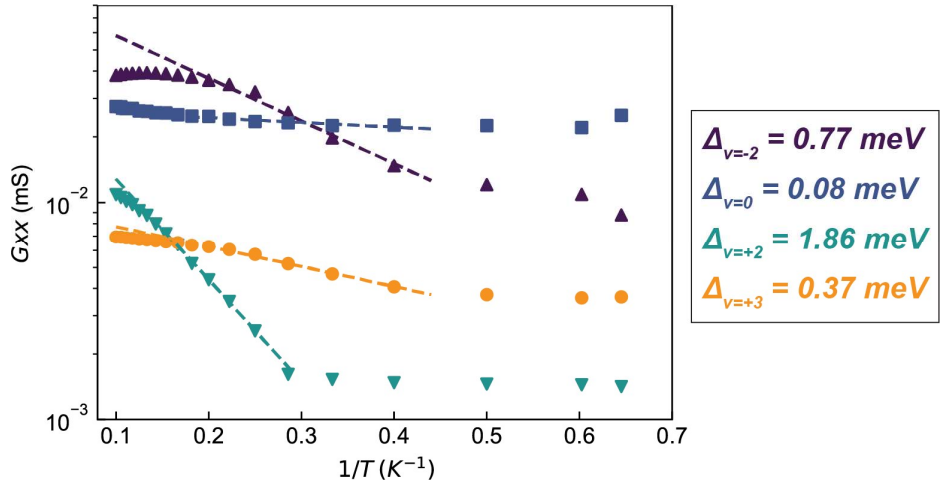

$\Delta_{\nu=-2} = 0.77$ meV
$\Delta_{\nu=0} = 0.08$ meV
$\Delta_{\nu=+2} = 1.86$ meV
$\Delta_{\nu=+3} = 0.37$ meV

**Figure S7. Thermal activation gaps for Device 1.** Arrhenius plot of the main correlated states at the integer fillings of Device 1.

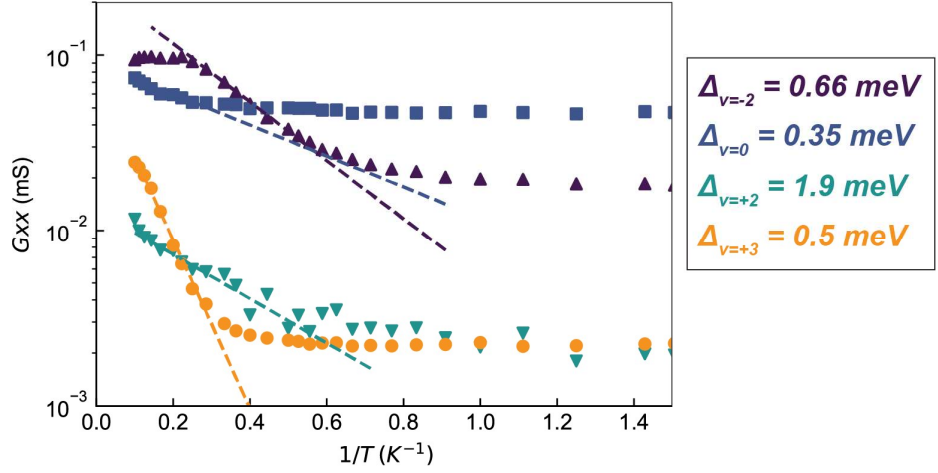

**Figure S8. Thermal activation gaps for Device 2.** Arrhenius plot of the main correlated states at the integer fillings of Device 2.

### d. Transport characteristics after top gate etching

We characterize the transport properties of the MATBG junctions after etching of the top gate. Figure S9a and S10a show the 4-terminal resistance across the junction as a function of the two top gates for Device 1 and Device 2, respectively. The vertical and horizontal resistive states corroborate the independent gate control of each side of the junction, as well as the presence of prominent correlated states after the last fabrication step. In Figures S9b and S10b we show the 4-terminal longitudinal resistance as a function of a single top gate ($V_{TG1}$) as the other one is set to charge-neutrality ($\nu_{TG2} = 0$). This is the experimental configuration used to measure the filling-dependent thermoelectric response $V_{PTE}(\nu_1)$ of the MATBG junctions in Figure 2 of the main text and onwards. We focus our comparison between the experimental data and theoretical calculations on the electron-doped flat band ($\nu_1 > 0$), as it exhibits stronger insulating states at integer fillings.

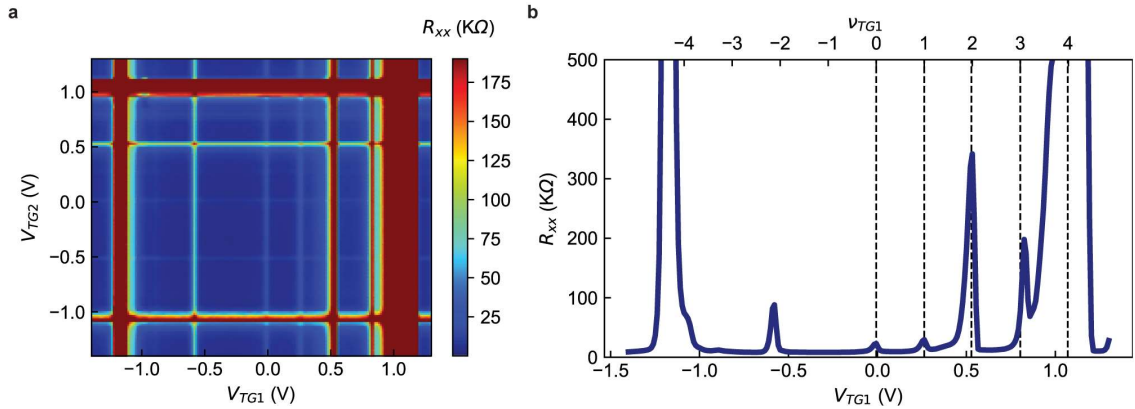

**Figure S9. Transport characteristics of Device 1 after top gate etching.** a) Dual-gate $R_{xx}(V_{TG1}, V_{TG2})$ map for Device 1 at $T_L = 35$ mK. b) 4-terminal $R_{xx}(V_{TG1})$ of Device 1 at $T_L = 35$ mK, $\nu_{TG2} = 0$.

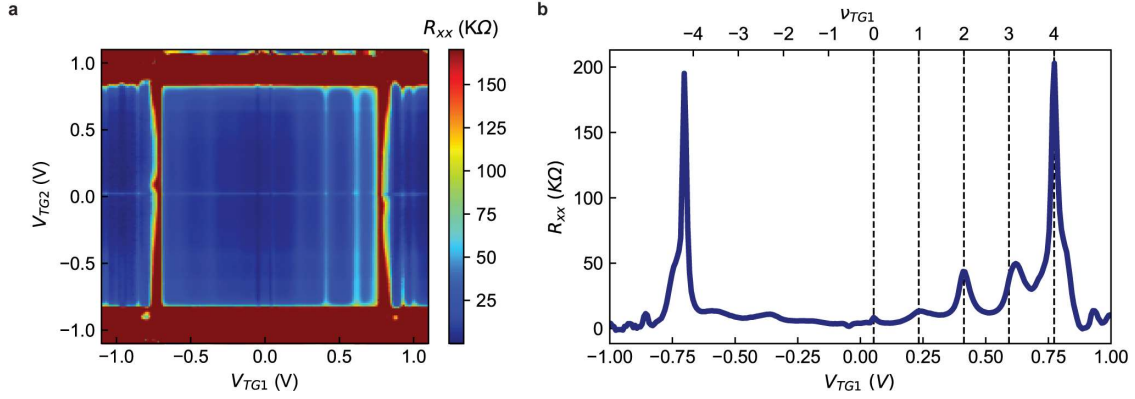

**Figure S10. Transport characteristics of Device 2 after top gate etching.** a) Dual-gate $R_{xx}(V_{TG1}, V_{TG2})$ map for Device 2 at $T_L = 6$ K. b) 4-terminal $R_{xx}(V_{TG1})$ of Device 2 at $T_L = 6$ K, $\nu_{TG} = 0$.

### III. Optoelectronic measurements

#### a. Optoelectronic setup

The optical setup for the experiment is schematically depicted in Figure S11a. We use a single-frequency, fiber-coupled CW laser operating at $\lambda = 1550$ nm (SFL1550P). The laser excitation is coupled to free space using a fiber-coupled collimating lens and is then mechanically modulated through the inner slot of a dual-slot chopper wheel. The input laser power is controlled using a fiber-coupled optical attenuator (JGR OA5). Scanning focused excitation of the sample is performed using a 4-$f$ scanning galvo microscope formed by XY galvanic mirrors, two spherical lenses and a 50x near-infrared objective lens (Olympus LMPLN50XIR) placed inside the cryostat. The reflected laser intensity is collected simultaneously with the scanning of the laser excitation. The scanning area is $\sim 60 \times 60$ μm$^2$. The sample position is adjusted using 3-axis piezo positioners. We measure the spot size at optimal focus using the knife-edge technique on a lithographically-defined metal contacts, finding a laser spot of $\sim 2$ μm in diameter. Lastly, the device is also confocally imaged with white light using a halogen lamp and a standard CMOS camera for visible wavelengths.

The thermoelectric response is read out using a Lock-in amplifier (SRS860) referenced to the local oscillator of the chopper wheel. We measure the voltage drop between the two contacts closest to the illuminated *pn*-junction (see Fig. S11b). All other contacts to the MATBG are floating. We use a low-noise voltage pre-amplifier with 6 dB band-pass filtering around the chopper frequency $(100 - 200$ Hz$)$. The two top gates are biased using DC sources (Yokogawa GS200) over $100$ MΩ resistors; while the silicon gate and graphite back gate remain grounded for all measurements in order to suppress cross-talk between top gates.

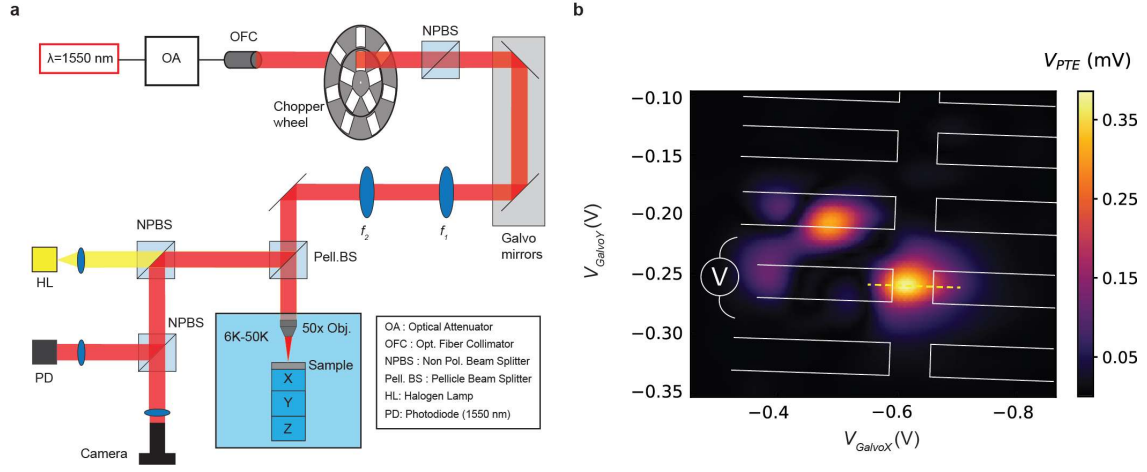

**Figure S11. Optoelectronic setup.** a) Schematic of the optoelectronic setup. b) Spatial photovoltage map of Device 2 at $T_L = 6$ K. The yellow lines show the Hall bar structure and the contacts used as voltage probes, obtained from simultaneous reflectivity scan.

### b. Photo-mixing measurements

We study the cooling dynamics of the MATBG *pn*-junctions using a CW photo-mixing technique[1–3]. The working principle and obtained results will be described in Section IV. As the modified optical setup for CW photo-mixing is not included in Figure 11a, we briefly describe it here.

A tunable CW laser with frequency $\lambda = 1550$ nm $\pm \Delta\lambda$ (Thorlabs TLX1) is mixed with the original fixed-frequency CW excitation. Another optical fiber collimator is free-space coupled using a fiber collimator. An acousto-optic modulator (AOM) is used to control its output intensity and the beam is then modulated through the outer slot of the dual-slot chopper wheel. Then, the laser beam is collinearly mixed with the other, fixed-frequency CW laser signal in a non-polarizing beam splitter and both follow the laser path described in Fig. S11a. The mixing of the laser fields produces an optical beating whose frequency can be controlled by tuning the wavelength of the second laser. The dual modulation scheme of the laser beams allows sensitive detection of the nonlinear response at the heterodyne difference frequency of the two chopping frequencies $f_{mix} = f_1 - f_2$.

### c. Photo-thermoelectric response

Dual top gate maps of the photo-thermoelectric response for each device are shown in Figure S12. Vertical and horizontal features appearing at integer fillings verify the independent control of each side of the *pn*-junction, as in the dual $R_{xx}$ maps in Fig. S9a, S10a. In the case of Device 2, the MATBG region under gate 1 (labelled as $\nu_1$) shows weaker correlated features at integer fillings. For the linecuts of $V_{PTE}$ vs. $\nu_2$ shown across the main text and SI, we fix $\nu_1 = 0$ and sweep the gate controlling the more-strongly correlated MATBG region.

We note that fine-tuning the voltage needed to have $S(v_1 = 0) = 0$ is not trivial; as $S(v)$ is a steep antisymmetric function around CNP. Therefore, we apply two different strategies to eliminate parasitic contributions from that side of the junction. One option is tracing the voltage applied to top gate $TG_1$ as we sweep $TG_2$; following the points of zero thermoelectric response. The other option is to measure $S(TG_2)$ for various values of $TG_1$ around CNP and compute the arithmetic mean of the resulting lineshapes to average out possible contributions from $v_1 = 0 \pm \delta$.

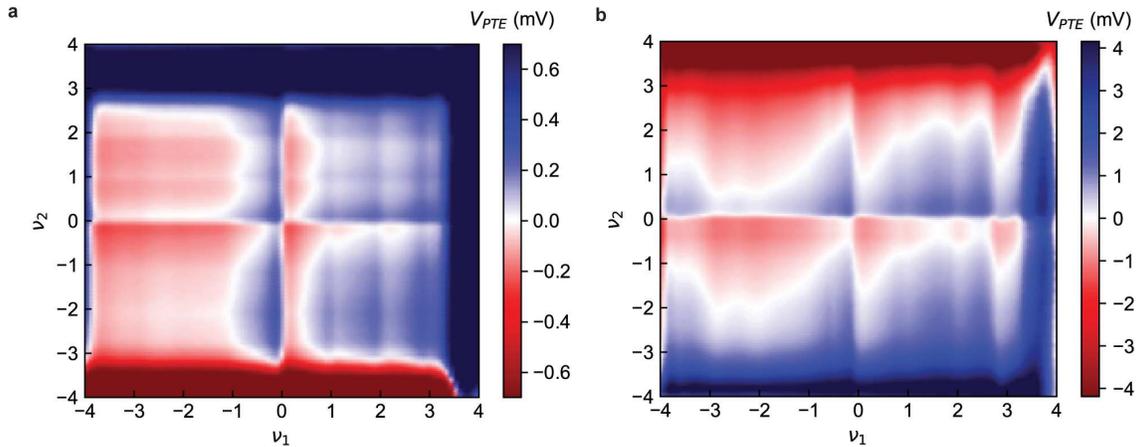

**Figure S12. PTE response of the devices across the flat bands.** a) Dual gate photovoltage map for Device 1 at $T_L = 10\ K$. b) Dual gate photovoltage map for Device 2 at $T_L = 10$ K.

As we approach full filling ($v_i = \pm 4$), a large photovoltage response appears. The origin of this response is likely connected to the larger gap between flat bands and dispersive bands. Thus, other photovoltage mechanisms besides the PTE effect might play a significant role. It is apparent from Figure S12 that the response from the band insulator 'leaks' into the flat bands and affects the thermoelectric response measured near $v = \pm 3$ (see $v_2 = \pm 3$, $v_2 = \pm 1$ and $v_1 = +3$ in Figure S12a and $v_2 = \pm 3$ in Figure S12b).

### IV. Thermal modelling and $\Delta T_e$ estimation

In this section, we corroborate the linear heating conditions in our experiment and estimate the electronic temperature increase $\Delta T_e$ induced by the CW laser input.

Based on the robust observation of 6-fold symmetry of the response in our samples (see Fig. S17 and main text), we can write the PTE response in the simple form $V_{PTE} = S\,\Delta T_e$[4,5]. In our experiment, the gate voltages control the Seebeck coefficient $S(v)$ through the carrier density, and the laser power controls the electronic temperature difference $\Delta T_e$ in the steady-state. Lattice temperature is fixed at $T_L = 10$ K in the following discussion.

As we do not perform *in situ* thermometry, a thermal model is required to make the connection between the absorbed laser power $P_{abs}$ and the induced electronic temperature increase $\Delta T_e$. We model the thermal response of the MATBG *pn*-junctions using a two-temperature model for a hot carrier system, where, in the steady-state under CW laser illumination, the electron gas will equilibrate at a higher temperature than the surrounding bath. Such a model is motivated by the well-known hot carrier effects that dominate the thermal response of graphene-like systems[4–7]. The electron gas of MATBG (with temperature $T_e$ and heat capacity $C_e$) exchanges heat with the environmental degrees of freedom (with temperature $T_L$ and heat capacity $C_L$) at a rate given by the thermal conductance $G_{th}$ between the two subsystems. We neglect the temperature increase of the environment as $C_L \gg C_e$ and the conversion efficiency between the absorbed optical power and electronic temperature increase is extremely efficient in graphene-like systems. Therefore, in the following we focus on the temperature of the electron gas and its laser-induced heating. We define $\Delta T_e = T_e'(P_{abs}) - T_e(P_{abs} = 0)$, where $T_e(P_{abs} = 0) = T_L$. Then, using the relation:

$$G_{th}(T_L) = \frac{C_e(T_L)}{\tau(T_L)} \qquad 1.1$$

the temperature increase of the electronic subsystem can be written as:

$$\Delta T_e(T_L) = \frac{P_{abs}}{G_{th}(T_L)} = \frac{P_{abs}\tau(T_L)}{C_e(T_L)} \qquad 1.2$$

where $\tau(T_L)$ is the thermal relaxation time for the electron subsystem. We assume that the absorbed power $P_{abs}$ does not depend significantly on $T_L$.

These equations, combined with the expression for the PTE voltage $V_{PTE} = S \cdot \Delta T_e$, set a relation between the excitation power $P_{abs}$, the laser-induced heating $\Delta T_e$, the Seebeck coefficient $S$ and the thermoelectric response $V_{PTE}$. We now inspect the power-dependence of the measured PTE response. An estimation for $\Delta T_e(T_L)$ based on the thermal properties of our samples is detailed in the next subsection.

      a. **Linear response at low excitation power**

Figure S13a (S14a) shows the power dependence of the PTE response near neutrality for Device 1 (Device 2). We focus here on the response near neutrality as it exhibits the clearest 6-fold symmetry and is generally not dependent on the strong electronic interactions and FS reconstructions that appear throughout the flat band. We observe a clear sublinear power-dependence, typical for thermal response mechanisms. The sublinear dependence arises due to the increase of heat capacity for increasing temperatures, supporting the thermoelectric origin of the photovoltage response. The transition between the linear and sublinear regimes is device and lattice temperature dependent.

Importantly, for low excitation powers the response evolves linearly with power, as $V_{PTE} \propto \Delta T_e \propto P_L$. We show in Figure S13b (S14b) how the power-normalized response $V_{PTE}/P_{abs}$ of Device 1 (Device 2) collapses into a single curve across the entire flat bands for low excitation powers, thus confirming low-heating conditions and a linear heating regime where $\Delta T_e \ll T_e$ for low laser powers. All in all, we corroborate that we realize linear heating conditions at low excitation powers, allowing us to use $V_{PTE}$ as an accurate probe for the evolution of the Seebeck coefficient $S(\nu)$ across the flat bands. We note here that other scenarios where $V_{PTE} \propto P_L$ without both quantities depending linearly on $\Delta T$ appear highly unlikely but cannot be discarded. Lastly, we comment that Device 1 exhibits a positive response near the electron-doped band insulator, where we would expect negative values. The sign of the response here is badly defined due to the massive resistance (~M$\Omega$) of this band insulator in Device 1, which forbids a faithful extraction of the phase of the signal for demodulation.

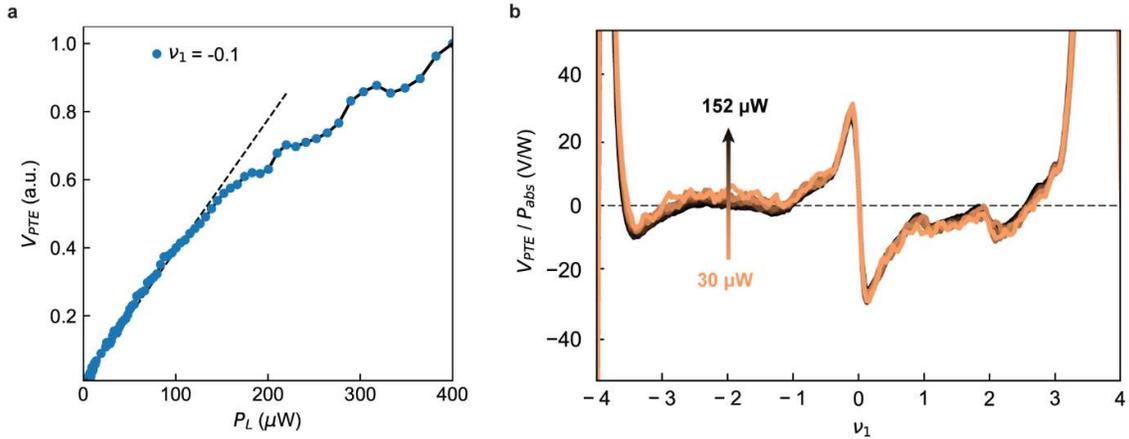

**Figure S13. Linear heating regime for Device 1.** a) Laser power dependence of the photo-thermoelectric voltage near neutrality for Device 1. $T_L = 10$ K. Color coding indicates the total incident power $P_L$ for each trace. b) Filling dependence of the thermoelectric response normalized by incident laser power in Device 1. $T_L = 10$ K.

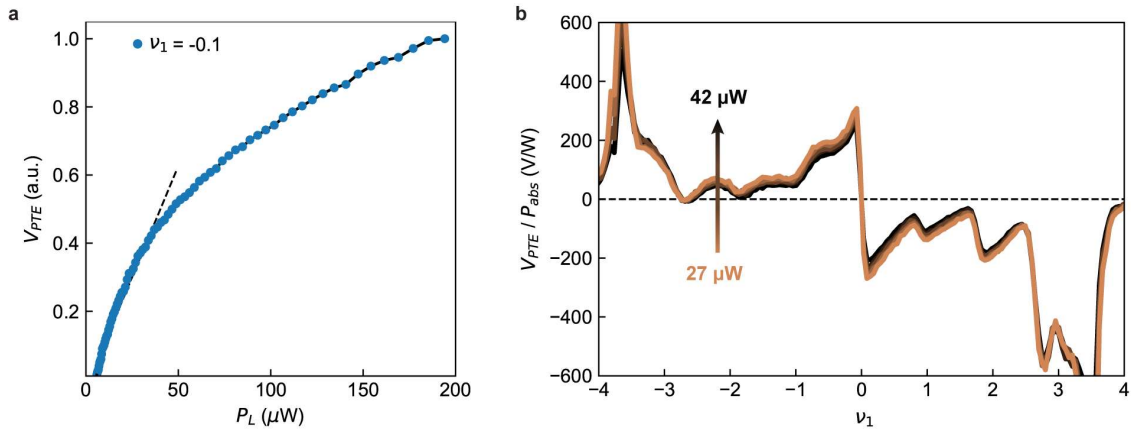

**Figure S14. Linear heating regime for Device 2.** a) Laser power dependence of the photo-thermoelectric voltage near neutrality for Device 2. $T_L = 10$ K. Color coding indicates the total incident power $P_L$ for each trace. b) Filling dependence of the thermoelectric response normalized by incident laser power in Device 2. $T_L = 10$ K.

### b. $\Delta T_e$ estimation

Having established the linearity of the response at low excitation powers, we now employ the two-temperature hot carrier model to estimate the temperature increase $\Delta T_e$ in our samples. In the following, we address the different terms in Eq. 1.2. both experimentally and theoretically in order to provide an estimate for $\Delta T_e(T_L)$.

#### b.1. Thermal relaxation time $\tau(T_L)$ of MATBG

The thermal relaxation time $\tau(T_L, \nu)$ in our MATBG *pn*-junctions is experimentally determined using a heterodyne CW photo-mixing technique[1–3]. This technique relies on the mixing of two CW laser beams with central wavelength $\lambda$, where one of them can be slightly detuned $\lambda_2 = 1550$ nm $\pm \Delta\lambda$ to create an optical beating of frequency $\Omega = \frac{2\pi}{c}(\frac{1}{\lambda_1} - \frac{1}{\lambda_2})$. The electronic temperature $T_e$ will then oscillate at the beating frequency $\Omega$. The amplitude of the oscillations will be damped for $\Omega > \tau^{-1}$ as the electron gas cannot relax to equilibrium within the oscillation cycle of the excitation, whereas the amplitude will be maximal for $\Omega \ll \tau^{-1}$. Using a double-modulation scheme (see Section III.b) we record the nonlinear response of the device $V_\Delta(P_1, P_2)$, where $P_i$ is the optical power from each laser, as a function of the optical beating frequency $\Omega$. $V_\Delta$ exhibits a Lorentzian dependence on $\Omega$, where $FWHM = (\pi\tau)^{-1}$. The thermal relaxation time measured here is agnostic to the cooling mechanisms at play, which could change under different experimental conditions[1].

The inset of Figure S15a shows a typical measurement of the filling-dependence of $\tau$ for Device 2. The thermal relaxation time along the flat bands is nearly constant, except near the band insulators where it increases up to $\tau \approx 20$ ps. Importantly, the thermal relaxation time $\tau$ also remains nearly-constant between $\tau \approx 3 - 5$ ps at the relevant temperature range. Figure S15b depicts two typical measurements of $\tau$ at bath temperatures of $T_L = 6$ K, 25 K, where we find $\tau = 3.30$ ps and $\tau = 4.98$ ps, respectively. In our calculations we use an average $\tau(T_L, \nu) \approx 4$ ps. The measurements presented in Fig.15b are performed away from the integer fillings. A detailed study of the cooling dynamics of MATBG samples can be found in previous work[8].

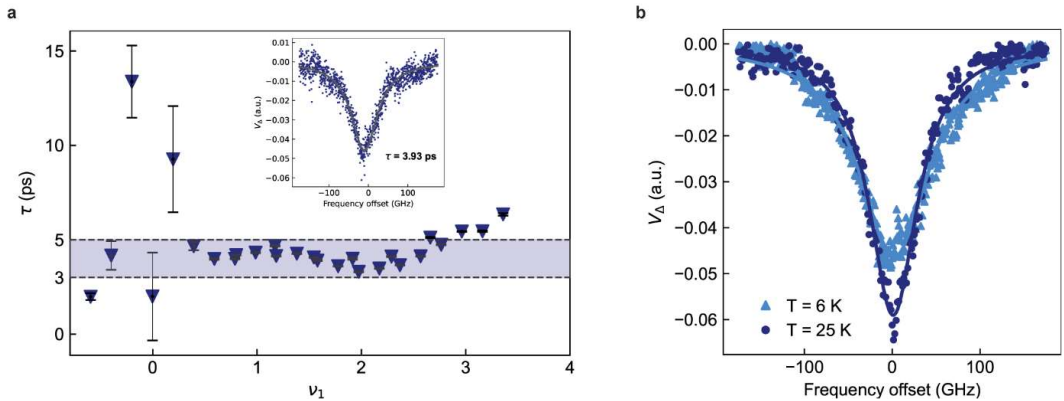

**Figure S15. Thermal relaxation time of MATBG *pn*-junctions.** a) Thermal relaxation time of Device 2 along the electron-doped flat bands. Inset shows a typical trace for a CW photo-mixing experiment. The FWHM of the Lorentzian lineshape is used to extract $\tau$. b) Thermal relaxation time of Device 2 at $T_L = 6$ K, 25 K. We consistently find $\tau \sim 3 - 5$ ps across the relevant temperature range.

### b.2. Absorption and coupling efficiency for MATBG *pn*-junctions

We calculate the absorbed optical power as $P_{abs} = P_{inc} \cdot \alpha \cdot \eta_{coupl}$, where $P_{inc}$, $\alpha$ and $\eta_{coupl}$ are the incident laser power, the absorption coefficient in the MATBG layers and the optical coupling efficiency. The incident laser power is calibrated using a power meter before each measurement run. To estimate the absorption coefficient, we simulate the electric field distribution inside the van der Waals heterostructure. Using an optical transfer matrix simulation[9], we calculate the absorption coefficient for the MATBG layers, finding $\alpha = 2.16\ \%$ for Device 1 and $\alpha = 3.33\ \%$ for Device 2.

The optical coupling efficiency is obtained as the spatial overlap between the Gaussian laser spot of diameter $\sim 2$ μm and the device dimensions (between the voltage probes highlighted in Fig. S1a). Introducing the device dimensions and the laser spot size into:

$$\eta_{coupl} = \frac{1}{2\pi\sigma^2} \int_{L/2}^{-L/2} \int_{W/2}^{-W/2} e^{-\frac{(x^2+y^2)}{2\sigma^2}}\, dxdy \qquad 2.2$$

where $\sigma$ is the laser spot radius, $L$ is the device length and $W$ is the device width. We find $\eta_{coupl} = 81.75\ \%$ for Device 1 and $\eta_{coupl} = 82.96\ \%$ for Device 2.

### b.3. Electronic heat capacity $C_e(T_L)$ of MATBG

In this section, we estimate the electronic heat capacity $C_e$ of our MATBG samples. We note that our goal here is not to present comprehensive temperature and filling-dependent results (which we leave for future work), but rather to provide an order of magnitude estimation of this quantity. To do so, we employ the THF model and compute its internal energy as a function of temperature and filling using second-order self-consistent perturbation theory in the symmetry-broken phases, as detailed in accompanying theoretical work[10]. In Figure S16 we present an exemplary computation for the specific heat of the KIVC state for $+1.8 < \nu < +2.5$ in the relevant temperature range $T_L \approx 0.6 - 0.7\ T_{ord}$. We find $C_p \approx 2 - 5\ x10^{-7} \frac{\text{J}}{\text{Km}^2}$ for a twist angle $\theta = 1.06°$. For our micrometer-scale samples, we find electronic heat capacities of $C_e \approx 1.4 - 3.5\ x10^{-1}\ \frac{\text{J}}{\text{K}}$ for Device 1 and $C_e \approx 0.82 - 2.06\ x10^{-1}\ \frac{\text{J}}{\text{K}}$ for Device 2.

We note that this estimation of $C_p$ is in good agreement with previous calculations for twisted bilayer graphene, considering the different lattice temperatures. Importantly, as we will show below, these values for the electronic heat capacity are consistent with the expected $\Delta T_e$ when comparing the computed THF Seebeck coefficient and the measured response $V_{PTE}$.

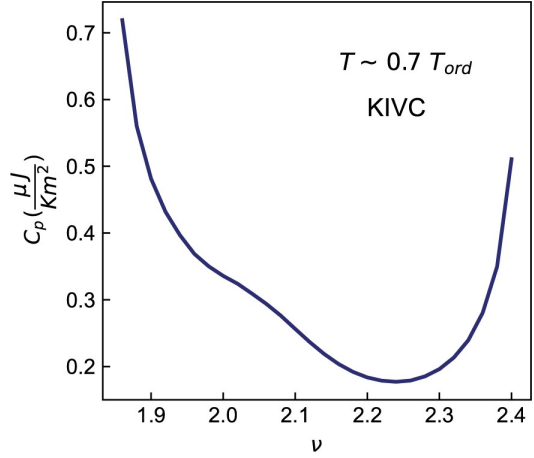

**Figure S16. Specific heat for a symmetry-broken ground state in the THF model.** Filling-dependent specific heat for an exemplary symmetry-broken ground state. We show here the specific case of the KIVC correlated insulator for $\nu = +2$, calculated using second-order self-consistent perturbation theory. This is obtained by computing the internal energy as a function of temperature and filling $U(T, \nu)$, as detailed in Ref. 10, and then computing $C_p = \frac{\partial U(T,\nu)}{\partial T}$.

### b.4. Estimating $\Delta T_e$

Lastly, we estimate the temperature increase $\Delta T_e$ for a representative low-power measurement. For Device 2 under 20 $\mu W$ incident optical power and $T_L = 10$ K, using Eq. 1.2. we obtain:

$$\Delta T_e = \frac{P_{abs} \cdot \tau}{C_e} = \frac{P_{inc} \cdot \alpha \cdot \eta_{coupl} \cdot \tau}{C_e} \approx \frac{20 \times 10^{-6} \text{ W} \cdot 0.033 \cdot 0.8296 \cdot 4 \times 10^{-12} \text{ s}}{2.06 \times 10^{-18} \frac{J}{K}} \approx 1.06 \text{ K}$$

The estimated temperature increase supports experimental evidence shown in Section IV.a; as well as the minimal assumptions in the thermal modelling of the MATBG p-n junctions. We can now use the estimated $\Delta T_e$ to make a quantitative comparison between the calculated Seebeck coefficient and the experimentally determined $V_{PTE}$. Our calculations find peak values for the Seebeck coefficient close to charge-neutrality of $S \approx 400$ μV/K. Taking the temperature increase of $\Delta T_e \approx 1.06$ K, we would obtain a response of $V_{PTE} \approx 425$ μV. Experimentally, we find $V_{PTE} \approx 300$ μV, which is in good agreement with the theoretical estimate. We note that the specific heat of MATBG has not been experimentally determined and depends strongly on multiple parameters such as doping, temperature and twist angle. All in all, our estimation for $\Delta T_e$ yields the right order of magnitude for the laser-induced heating and validates the use of the linear heating approximation.

It is worth noting that the $\Delta T_e$ acts as an upper bound for the actual temperature increase in the experiment. In the model, we assumed that no energy is deposited on the phononic degrees of freedom and the efficiency in the energy-to-temperature conversion for the carriers in MATBG is 100 %. Optical phonon emission from very high temperature carriers can reduce the actual heat deposited on the electron subsystem. Experimentally, we observe clear signatures of PTE voltage generation at much higher laser powers (see Figs. S17 and S18) where non-linear effects could play a large role; even in the highly non-equilibrium conditions of pump-probe measurements[8]. All in all, the description of the voltage generation in terms of Seebeck-driven thermoelectric transport in the linear heating approximation seems to hold for higher laser powers than those from theory.

Lastly, we note that for both devices the linear heating approximation remains valid up to temperature increases of $\Delta T_e \sim 2 - 3$ K for a lattice temperature of $T_L = 10$ K. The disparity in the incident powers disparity for sublinear response (Fig. S13, S14) stems from differences between the two samples. Device 1 has a lower absorption coefficient and larger device dimensions, which lead to an overall higher electronic heat capacity.

### V. Photovoltage-generation mechanisms

In this section we discuss the physical picture for the photo-thermoelectric voltage generation in MATBG *pn*-junctions. We also discuss and provide evidence for the absence of other photovoltage generation mechanisms in our experiment.

#### a. Light matter interaction in MATBG for near-infrared illumination

As the energy of the incident telecom photons ($\lambda = 1550$ nm, $h\nu = 0.8$ eV) is much larger than the energy scale of the flat bands ($\sim 10 - 50$ meV), the dynamics of light-matter interaction at this wavelength resemble those of untwisted bilayer graphene. As put forward in previous reports[11,12] we consider the dominant absorption mechanism to be the inter-band transitions between the dispersive valence and conduction bands.

After absorption, few electrons (holes) with very high temperature share their energy through *e-e* scattering on a femtosecond scale until reaching the bottom (top) of the conduction (valence) band. Then, the hot carriers relax into the MATBG flat bands and thermalize into a new Fermi-Dirac distribution of higher temperature $T_L + \Delta T_e$. We note here that these dynamics have not extensively been studied yet for MATBG, but the observation of ultrafast electron cooling down to the lowest temperatures[8] (up to an order of magnitude faster than non-moiré bilayer graphene at $T_L = 5$ K[9,13,14]) suggests that the dynamics of carrier heating, scattering and eventual thermalization are not slower than those of graphene.

This picture, in which photo-absorption of near-infrared photons in MATBG is largely insensitive to the detailed electronic structure of its flat bands, has been experimentally backed through direct measurements of infrared photoresistance[12] and indirect studies of the system's photoluminesce[13–15]. Indeed, these experiments point towards a hot carrier distribution populating the flat band populated shortly after infrared excitation.

#### b. Photo-thermoelectric effect as dominant photovoltage mechanism

In this work, we provide clear evidence for photovoltage generation via the photo-thermoelectric effect in MATBG *pn*-junctions, analogous to the observations of PTE effect in high-quality gate-defined graphene *pn*-junctions[5,16]. In Figure S17 we highlight the observation of the hallmark 6-fold symmetry[4,5] of the photovoltage signal around charge-neutrality for several MATBG devices and a relaxed Bernal bilayer graphene.

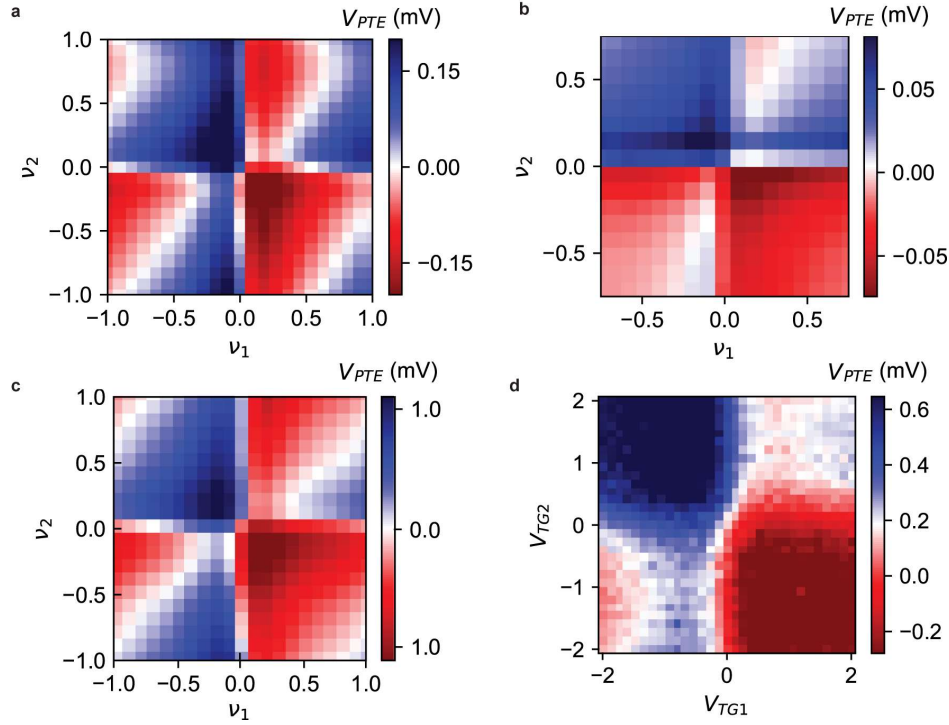

**Figure S17. 6-fold symmetry of the photoresponse near charge-neutrality for several devices.** a) Device 1 (1.14º), $T_L = 10$ K, $P_L = 120$ µW. b) Device 2 (1.06º), $T_L = 6$ K, $P_L = 3$ µW. c) Device 3 (1.24º), $T_L = 7$ K, $P_L = 1.25$ mW. d) Device 4 (0º), $T_L = 300$ K, $P_L = 50$ µW.

We also note that the 6-fold symmetry also appears for a wide range of lattice temperatures and incident optical powers across the different samples. The robustness of the symmetry of the response across devices and experimental conditions, combined with the re-appearance of the 6-fold symmetry of the response at integer fillings provides solid evidence for the PTE effect as the dominant mechanism for photovoltage generation in our experiment. Lastly, the sublinear power dependence of the response (Fig. S13a, S14a) is indicative of a thermal mechanism, as discussed in Section IV.a. This sublinear behavior is observed across multiple samples and experimental conditions, including pump-probe studies where the electron temperature is orders of magnitude larger than in the measurements presented here.

Thermoelectric transport can be generally described by the Mott relations:

$$\sigma(\mu) = \int_{-\infty}^{\infty} dE \left(-\frac{\partial n_F(E)}{\partial E}\right) \sigma(\mu, E) \approx \sigma(\mu, 0)$$

$$\sigma(\mu)S(\mu) = -\frac{1}{eT}\int_{-\infty}^{\infty} dE \left(-\frac{\partial n_F(E)}{\partial E}\right) E\, \sigma(\mu, E) \approx -\frac{\pi^2 T}{3e} \frac{\partial \sigma(\mu, E)}{\partial E}\bigg|_{E=0}$$

where $\sigma(\mu)$ and $S(\mu)$ are the electrical conductivity and Seebeck coefficient at chemical potential $\mu$, $e$ is the electron charge, $T$ is the temperature, $n_F(E)$ is the Fermi-Dirac distribution and $\sigma(\mu, E)$ is the energy-dependent conductivity function. The last approximations correspond to the Sommerfeld expansion at low temperatures. Phenomenologically, $\sigma(\mu, E)$ depends on the density of states $N$, the carrier lifetime $\tau$ and the carrier effective mass $m^*$ as $\sigma \propto N\tau/m^*$.

The low-temperature Seebeck coefficient of the MATBG flat bands was found to deviate from the semiclassical Mott formula[25–28], given by $S_{Mott} = -\frac{\pi^2}{3ek_BT}\frac{\partial}{\partial \mu}\log \sigma(\mu)$, near integer fillings $\nu$. Notably, the expression for $S_{Mott}$, unlike the Mott relations, relies on the 'rigid band' approximation $\sigma(\mu, E) \approx \sigma(\mu + E, 0)$, enabling the use of the experimentally measurable conductivity $\sigma(\mu)$. However, the 'rigid band' approximation cannot be applied to the highly interacting flat bands of MATBG. Hence, comprehensive modelling of the thermoelectric transport in the flat bands must account for the effects of electron interactions in the energy-dependent conductivity function $\sigma(\mu, E)$.

### c. Other photovoltage mechanisms

Here we discuss other plausible photovoltage generation mechanisms. As the *pn*-junctions are operated under zero applied bias, the bolometric and photoconductive effects can be readily discarded. Photovoltage arising from resonant plasma waves in our sample can also be neglected as we do not utilize any plasmonic resonant structure or tailor our device's size or shape for this purpose. Furthermore, similar response is observed for samples with different size, form factors and microscopic details such as the twist angle or the thickness of different components of the vdW heterostructure.

We find no evidence for photogating effects in the studied experimental conditions. The position and width of the integer-filling oscillations in $V_{PTE}$ do not change appreciably with increasing laser power (Fig. S18); indicating the absence of appreciable photogating effects in our experiment. Furthermore, previous reports of the infrared photoresistance of MATBG did not observe any shift between the peak positions of longitudinal resistance and photoresistance[12].

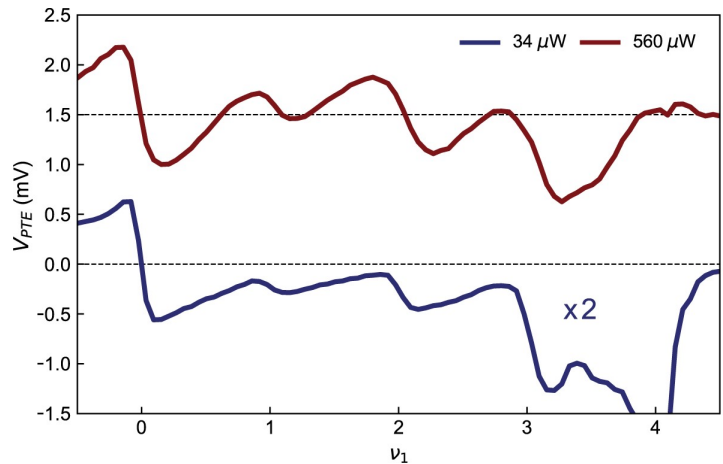

**Figure S18. Absence of photogating effect.** PTE response at $T_L = 10$ K for low and high incident powers. The oscillatory features arising from the correlated ground states are not shifted appreciably with increasing powers. The trace for 560 µW is vertically shifted by 1.5 mV for clarity and the magnitude of the 34 µW has been doubled.

Several factors point towards negligible photovoltaic (PV) contributions to the measured photovoltage. The main evidence against the PV effect is the already discussed 6-fold symmetry of the response around charge-neutrality[4,5]. In the scenario of a PV-dominated response, the sign of the generated photovoltage depends only on the polarity of the junction and yields a 2-fold symmetry of the response near CNP which is clearly not present in our data. The additional sign changes unequivocally reflect the non-monotonic Seebeck coefficient of Dirac electrons as the driving force of the measured response[16]. Furthermore, PV contributions are expected to be most pronounced near CNP where the lower carrier density leads to reduced *e-e* scattering events that quench the PV response[4]. The ultrafast nature of the response[8] strongly suggests the hot-carrier nature of the response, in disagreement with a PV scenario.

We do not observe saturation of the device response with optical power (see Figures S13a and S14a), as could be expected in a PV mechanism. We also point out that the laser excitation is focused at the center of the pn-junction, away from any metallic contact (see Fig. S11b), which are known to generate PV contributions to the optoelectronic response of graphene[17,18]. In addition, we note that in the case of non-moiré graphene, the dominance of the PTE contributions over the PV response has been well established both theoretically and experimentally[4,5,16].

Here, we discuss the contributions to thermoelectric transport from the phonon drag effect. This contribution, which is additive to the diffusion thermopower discussed up to this point, arises due to the heat conduction through the lattice phonons which then 'drag' the charge carriers towards the colder parts of the device[19,20]. Quantifying the expected magnitude of the phonon-drag thermopower in TBG is challenging as neither the complete phononic band structure nor the *e-ph* coupling strength of the system are well understood so far. However, we posit that the phonon drag effect does not play a significant role in our measurement based on the following observations.

First, the optical excitation selectively heats up the electronic subsystem, as discussed before and the electronic heat capacity $C_e$ is far smaller than the heat capacity of the (phononic) bath $C_L$. Hence, at equilibrium, the system has hot carriers interacting with a comparatively cold phonon bath, from which significant phonon drag signal is not expected.

Second, the phonon drag thermopower is enhanced in the presence of high-mobility phonons and low-mobility, large-mass carriers[21]. At first order, we expect the lowest mobility for carriers in the flat bands near charge-neutrality. However, as already established before, the observed response around CNP is naturally explained by the PTE effect. Therefore, we do not expect significant phonon drag contributions away from CNP as the carrier density and mobility increase. Lastly, we note that previous reports of the in-plane thermoelectricity of MATBG have not found significant phonon drag contributions[22,23]. In order to shed light on the phonon drag effect in MATBG and its contributions to thermoelectric transport, experiments analogous to those realized in bulk materials ought to be explored.

## VI. Theoretical modelling

The thermoelectric effect of twisted bilayer graphene is modeled microscopically as explained in the accompanying works[10,24]. The full details in addition to further numerical results are relegated therein. Here we provide only a short overview of the methodology employed. The THF model parameters used in this work are listed in said papers and correspond to a twist angle $\theta = 1.06°$ and a double-gated setup with a distance between gates of $\xi = 10$ nm, in accordance with Device 2 in this work. For both the interacting and non-interacting results presented here, we derive the expression for the Seebeck coefficient in terms of the Green's function of the system, using linear response theory[24].

For the results on the symmetry-broken states presented in Figure 3 of the main text, the interacting Green's function of the system is obtained using second-order self-consistent perturbation theory[10]. Within this formalism, all the terms of the interacting THF Hamiltonian are treated up to at least first-order in perturbation theory (i.e. at least at the Hartree-Fock level). To account for correlation effects, for the leading-interaction term (the on-site Hubbard repulsion between f-electrons), we include the second-order perturbative correction[10]. The self-consistent simulations are performed starting from the zero temperature, integer-filled solution and incrementally doping the system up to $\Delta\nu = \pm 0.5$.

Because the self-consistent solution we obtain numerically for the $\nu = +2$ KIVC state does not feature an indirect gap, the corresponding Seebeck coefficient (which is nevertheless negative) has a smaller magnitude compare to the one inferred from the experimental data[24]. To better reproduce the experimental thermoelectric response, we 'artificially' increase the gap by increasing the $W_1$ and $W_3$ parameters of the THF model[25] by 12.5%. We note that this modification only changes the magnitude of the Seebeck coefficient, but not its shape or sign.

In the symmetric state (whose theoretically computed Seebeck coefficient is shown in Figure 4c of the main text), we obtain the interacting Green's function of the system by combining the Hartree-Fock and dynamical mean field theories[10]. Specifically, the entire THF interaction Hamiltonian is treated at the Hartree-Fock level, while the dynamic correlation effects stemming from the on-site Hubbard repulsion of f-electrons are obtained using dynamical mean field theory. We use a modified iterative perturbation theory solver for the latter, which is detailed and discussed at length in accompanying theoretical work[10].

In the non-interacting case (shown by the light gray lines in Figs. 4a, 4c of the main text), we employ $\Gamma_c = 1$ meV and $\Gamma_f = 4$ meV as the lifetime broadening factors for the c- and f-electrons. An increased broadening factor is added to the f-electrons in the non-interacting case in order to match the offset of the Seebeck coefficient in the symmetric state, wherein the f-electrons experience a large interaction-driven scattering rate. Lastly, for the two-band model presented in Figure 2, derivations, additional numerical results at different temperatures, as well as analytical results (including asymptotic expressions for the Seebeck coefficient arising in various limits) are presented in accompanying theoretical work[24].